\documentclass[twocolumn,superscriptaddress,showpacs]{revtex4}

\usepackage{bm}

\begin{document}

\title{Statistical mechanics of 2D turbulence with a prior vorticity distribution}

\author{Pierre-Henri Chavanis}
\email{chavanis@irsamc.ups-tlse.fr}
\affiliation{ Laboratoire de Physique Th\'eorique, Universit\'e Paul
Sabatier, 118 route de Narbonne 31062 Toulouse, France}


\begin{abstract}
We adapt the formalism of the statistical theory of 2D turbulence in
the case where the Casimir constraints are replaced by the
specification of a prior vorticity distribution. A phenomenological
relaxation equation is obtained for the evolution of the
coarse-grained vorticity. This equation monotonically increases a
generalized entropic functional (determined by the prior) while
conserving circulation and energy. It can be used as a thermodynamical
parametrization of forced 2D turbulence, or as a numerical algorithm
to construct (i) arbitrary statistical equilibrium states in the sense
of Ellis-Haven-Turkington (ii) particular statistical equilibrium
states in the sense of Miller-Robert-Sommeria (iii) arbitrary
stationary solutions of the 2D Euler equation that are formally
nonlinearly dynamically stable according to the Ellis-Haven-Turkington
stability criterion refining the Arnold theorems.
\end{abstract}



\maketitle


\section{\label{intro}Introduction}

Two-dimensional incompressible and inviscid flows are
described by the 2D Euler equations
\begin{equation}
{\partial \omega\over\partial t}+{\bf u}\cdot \nabla \omega=0,\qquad \omega=-\Delta\psi,\qquad {\bf u}=-{\bf z}\times\nabla\psi,
\label{intro1}
\end{equation}
where $\omega$ is the vorticity and $\psi$ the streamfunction.  The 2D
Euler equations are known to develop a complicated mixing process
which ultimately leads to the emergence of large-scale coherent
structures like jets and vortices. Jovian atmosphere shows a wide
diversity of structures: Jupiter's great red spot, white ovals, brown
barges,... One question of fundamental interest is to understand and
predict the structure and the stability of these quasi stationary
states (QSS). To that purpose, Miller \cite{miller} and Robert \&
Sommeria
\cite{rs} have proposed a statistical mechanics of the 2D Euler
equation (a similar statistical theory had been developed earlier by
Lynden-Bell \cite{lb} to describe the violent relaxation of collisionless
stellar systems governed by the Vlasov equation; see \cite{csr} for a
description of this analogy). The key idea is to replace the
deterministic description of the flow $\omega({\bf r},t)$ by a
probabilistic description where $\rho({\bf r},\sigma,t)$ gives the
density probability of finding the vorticity level $\omega=\sigma$ in
${\bf r}$ at time $t$. The observed (coarse-grained) vorticity field
is then expressed as $\overline{\omega}({\bf r},t)=\int \rho\sigma
d\sigma$. To apply the statistical theory, one must first specify the
constraints attached to the 2D Euler equation. The circulation
$\Gamma=\int
\overline{\omega}d{\bf r}$ and the energy $E={1\over 2}\int
\overline{\omega}\psi d{\bf r}$ will be called {\it robust
constraints} because they can be expressed in terms of the
coarse-grained field $\overline{\omega}$ (the energy of the
fluctuations can be neglected). These integrals can be
calculated at any time from the coarse-grained field
$\overline{\omega}({\bf r},t)$ and they are conserved by the dynamics. By contrast, the Casimir invariants $I_{f}=\int
\overline{f(\omega)}d{\bf r}$, or equivalently the fine-grained
moments of the vorticity $\Gamma_{n>1}^{f.g.}=\int
\overline{\omega^{n}}d{\bf r}$,
where $\overline{\omega^{n}}=\int \rho\sigma^{n}d\sigma$, will be
called {\it fragile constraints} because they must be expressed in
terms of the fine-grained vorticity. Indeed, the moments of the
coarse-grained vorticity $\Gamma_{n>1}^{c.g}=\int
\overline{\omega}^{n}d{\bf r}$ are not conserved since
$\overline{\omega^{n}}\neq \overline{\omega}^{n}$ (part of the
coarse-grained moments goes into fine-grained
fluctuations). Therefore, the moments $\Gamma_{n>1}^{f.g.}$ must be
calculated from the fine-grained field $\omega({\bf r},t)$ or from the
initial conditions, i.e. before the vorticity has mixed. Since we
often do not know the initial conditions nor the fine-grained field,
the Casimir invariants often appear as ``hidden constraints''.

The statistical theory of Miller-Robert-Sommeria (MRS) is based on
three assumptions: (i) it is assumed that the evolution of the flow
is strictly described by the 2D Euler equation (no forcing  and no
dissipation); (ii) it is assumed that we know the initial conditions
(or equivalently the value of all the Casimirs) in detail; (iii) it
is assumed that mixing is efficient and that the evolution is
ergodic so that the system will reach, at statistical equilibrium,
the most probable (most mixed) state. Within these
assumptions \footnote{Some attempts have been proposed to go beyond
the assumptions of the statistical theory. For example, Chavanis \&
Sommeria \cite{jfm2} consider a {\it strong mixing limit} in which
only the first moments of the vorticity are relevant instead of the
whole set of Casimirs. They also introduce the concept of {\it
maximum entropy bubbles} (or restricted equilibrium states) in order
to account for situations where the evolution of the flow is not
ergodic in the whole available domain but only in a subdomain.}, the
statistical equilibrium state of the 2D Euler equation is obtained
by maximizing the mixing entropy
\begin{equation}
\label{intro2}
S\lbrack \rho\rbrack=-\int \rho\ln\rho \ d{\bf r}d\sigma,
\end{equation}
at fixed energy $E$ and circulation $\Gamma$ (robust constraints) and
fixed fine-grained moments $\Gamma_{n>1}^{f.g.}$ (fragile
constraints). We must also account for the normalization condition
$\int\rho d\sigma=1$. This optimization principle is solved by
introducing Lagrange multipliers, writing the first order variations
as \cite{rs,jfm2}:
\begin{equation}
\label{intro3}
\delta S-\beta\delta E-\alpha\delta\Gamma-\sum_{n>1}\alpha_{n}\delta\Gamma^{f.g.}_{n}-\int \zeta({\bf r})\delta\rho d\sigma d{\bf r}=0.
\end{equation}

In the MRS approach, the conservation of all the Casimirs has to be
taken into account.  However, in geophysical situations, the flows are
forced and dissipated at small scales (due to convection in the jovian
atmosphere) so that the conservation of the Casimirs is
destroyed. Ellis-Haven-Turkington \cite{ellis} have proposed to treat
these situations by fixing the conjugate variables $\alpha_{n>1}$
instead of the fragile moments $\Gamma_{n>1}^{f.g.}$ (this is
essentially a suggestion that has to be tested in practice).  If we
view the vorticity levels as species of particles, this is similar to
fixing the chemical potentials instead of the total number of
particles in each species. Therefore, the idea is to treat the fragile
constraints {\it canonically}, whereas the robust constraints are
still treated {\it microcanonically}. A rigorous mathematical
formalism has been developed in \cite{jsp} and a more physical
presentation has been given in \cite{physicaD}. In the EHT approach,
the relevant thermodynamical potential (grand entropy) is obtained
from the mixing entropy (\ref{intro2}) by using a Legendre transform
with respect to the fragile moments \cite{physicaD}:
\begin{equation}
\label{intro4}
S_{\chi}= S-\sum_{n>1}\alpha_{n}\ \Gamma^{f.g.}_{n}.
\end{equation}
Expliciting the fine-grained moments, we obtain the relative (or grand) entropy
\begin{equation}
\label{intro5}
S_{\chi}\lbrack \rho\rbrack=-\int \rho\ \ln\biggl\lbrack {\rho\over \chi(\sigma)}\biggr\rbrack \ d{\bf r}d\sigma,
\end{equation}
where we have defined the {\it prior vorticity distribution}
$\chi(\sigma)\equiv{\rm exp}\lbrace
-\sum_{n>1}\alpha_{n}\sigma^{n}\rbrace$. We shall assume that this
function is imposed by the small-scale forcing so it has to be given
{\it a priori} as an input in the theory \cite{ellis,jsp,physicaD}.

\section{\label{e}Equilibrium statistical mechanics with a prior vorticity distribution}

When a prior vorticity distribution is given, the statistical
equilibrium state is obtained by maximizing the relative (or grand) entropy
$S_{\chi}$ at fixed energy $E$, circulation $\Gamma$ and normalization
condition $\int\rho d\sigma=1$ (grand microcanonical ensemble).  The
conservation of the Casimirs has been replaced by the specification of
the prior $\chi(\sigma)$. Writing the first order variations as
$\delta S_{\chi}-\beta\delta E-\alpha\delta\Gamma-\int \zeta({\bf
r})\delta\rho d\sigma d{\bf r}=0$, we get the Gibbs state
\begin{equation}
\label{e1}
\rho({\bf r},\sigma)={1\over Z({\bf r})}\chi(\sigma)  e^{-(\beta\psi+\alpha)\sigma},
\end{equation}
with $ Z=\int_{-\infty}^{+\infty} \chi(\sigma)
e^{-(\beta\psi+\alpha)\sigma}d\sigma$. This is the product of a
universal Boltzmann factor by a non-universal function $\chi(\sigma)$
fixed by the forcing. The coarse-grained vorticity is given by
\begin{equation}
\label{e2}
\overline{\omega}={\int \chi(\sigma)\sigma e^{-(\beta\psi+\alpha)\sigma} d\sigma\over \int \chi(\sigma) e^{-(\beta\psi+\alpha)\sigma} d\sigma}=F(\beta\psi+\alpha),
\end{equation}
with $F(\Phi)=-(\ln\hat\chi)'(\Phi)$, where we have defined
$\hat{\chi}(\Phi)=\int_{-\infty}^{+\infty}\chi(\sigma)
e^{-\sigma\Phi}d\sigma$. It is easy to show that
$F'(\Phi)=-\omega_{2}(\Phi)\le 0$, where
$\omega_{2}=\overline{\omega^{2}}-\overline{\omega}^{2}\ge 0$ is the
local centered variance of the vorticity. Therefore, $F(\Phi)$ is a
decreasing function.  Since $\overline{\omega}=f(\psi)$, the
statistical theory predicts that the coarse-grained vorticity
$\overline{\omega}({\bf r})$ is a {\it stationary solution} of the 2D
Euler equation and that the $\overline{\omega}-\psi$ relationship is a
{\it monotonic} function which is increasing at {negative
temperatures} $\beta<0$ and decreasing at positive temperatures
$\beta>0$. We have $\overline{\omega}'(\psi)=-\beta\omega_{2}$.  We
note that the $\overline{\omega}-\psi$ relationship predicted by the
statistical theory can take a wide diversity of forms
(usually non-Boltzmannian) depending on the prior $\chi(\sigma)$. Furthermore,
the coarse-grained distribution (\ref{e2}) extremizes a generalized
entropy in $\overline{\omega}$-space of the form \cite{pre}:
\begin{equation}
\label{e3}
S\lbrack \overline{\omega}\rbrack =-\int C(\overline{\omega}) d{\bf
r},
\end{equation}
at fixed circulation and energy (robust constraints). Writing the
first order variations as $\delta S-\beta\delta E-\alpha\delta
\Gamma=0$, leading to
\begin{equation}
\label{e4}
C'(\overline{\omega})=-\beta\psi-\alpha,
\end{equation}
and comparing with Eq. (\ref{e2}), we find that
$C'(x)=-F^{-1}(x)$. Therefore, $C$ is a convex function $(C''>0$)
determined by the prior $\chi(\sigma)$ encoding the small-scale
forcing according to the relation
\begin{equation}
\label{e5} C(\overline{\omega})=-\int^{\overline{\omega}}F^{-1}(x)dx=-\int^{\overline{\omega}}\lbrack (\ln {\hat \chi})'\rbrack^{-1}(-x)dx.
\end{equation}
We have
$\overline{\omega}'(\psi)=-\beta/C''(\overline{\omega})$.  Comparing with $\overline{\omega}'(\psi)=-\beta\omega_{2}$  we find that, at statistical equilibrium
\begin{equation}
\omega_{2}={1/ C''(\overline{\omega})}, \label{e6}
\end{equation}
which links the centered variance of the vorticity to the
coarse-grained vorticity and the generalized entropy. It also clearly
establishes that $C''>0$. On the other hand, the equilibrium
coarse-grained vorticity $\overline{\omega}({\bf r})$ {maximizes} the
generalized entropy (\ref{e3})-(\ref{e5}) at fixed circulation and
energy iff $\rho({\bf r},\sigma)$ maximizes $S_{\chi}$ at fixed $E$,
$\Gamma$ (see Appendix \ref{sec_ge} and \cite{ellis,bouchet}).  {\it
Therefore, the maximization of $S[\overline{\omega}]$ at fixed $E$ and
$\Gamma$ is a necessary and sufficient condition of EHT
thermodynamical stability.}

The preceding relations are also valid in the MRS approach except that
$\chi(\sigma)$ is determined {\it a posteriori} from the initial
conditions by relating the Lagrange multipliers $\alpha_{n>1}$ to the
Casimir constraints $\Gamma^{f.g.}_{n>1}$. In this case of freely
evolving flows, the generalized entropy (\ref{e3})-(\ref{e5}) depends
on the initial conditions, while in the case of forced flows
considered here, it is intrinsically fixed by the prior vorticity
distribution. On the other hand, a maximum of $S_{\chi}[\rho]$ at fixed
$E$ and $\Gamma$ is always a maximum of $S[\rho]$ at fixed $E$,
$\Gamma$ and $\Gamma_{n>1}^{f.g.}$ (more constrained
problem).  Therefore, a maximum of the generalized entropy
$S[\overline{\omega}]$ at fixed $E$ and $\Gamma$ determines a
statistical equilibrium state in the MRS viewpoint
\cite{bouchet}. However, the converse is wrong in case of ``ensemble
inequivalence'' \cite{ineq,bb} with respect to the conjugate variables
$(\Gamma_{n>1}^{f.g.},\alpha_{n})$ because a maximum of $S[\rho]$ at
fixed $E$, $\Gamma$ and $\Gamma_{n>1}^{f.g.}$ is not necessarily a
maximum of $S_{\chi}[\rho]$ at fixed $E$ and $\Gamma$. {\it Therefore, the
maximization of $S[\overline{\omega}]$ at fixed $E$ and $\Gamma$ is
a sufficient (but not necessary) condition of MRS thermodynamical
stability.}

\section{\label{relax}Relaxation towards equilibrium}

In the case where a small-scale forcing imposes a prior vorticity
distribution $\chi(\sigma)$, it is possible to propose a
thermodynamical parametrization of the turbulent flow in the form of
a relaxation equation that conserves the circulation and the energy
(robust constraints) and that increases the generalized entropy
(\ref{e3})-(\ref{e5}) fixed by the prior. This equation can be
obtained from a generalized Maximum Entropy Production principle
(MEPP) in $\overline{\omega}$-space \cite{pre}. We write
$\omega=\overline{\omega}+\tilde\omega$ and take the local average
of the 2D Euler equation (\ref{intro1}). This yields
$D\overline{\omega}/Dt=-\nabla\cdot \overline{\tilde\omega\tilde
{\bf u}}\equiv -\nabla\cdot {\bf J}$ where $D/Dt=\partial/\partial
t+\overline{\bf u}\cdot\nabla$ is the material derivative and ${\bf
J}$ is the turbulent current. Then, we determine the optimal current
${\bf J}$ which maximizes the rate of entropy production $\dot
S=-\int C''(\overline{\omega}){\bf J}\cdot
\nabla\overline{\omega}d{\bf r}$ at fixed energy $\dot E=\int {\bf
J}\cdot \nabla\psi d{\bf r}=0$, assuming that the energy of the
fluctuations ${\bf J}^{2}/2\overline{\omega}$ is bounded. According
to this phenomenological principle, we find that the coarse-grained
vorticity evolves according to \cite{pre,physicaD}:
\begin{equation}
{\partial \overline{\omega}\over\partial t}+{\bf u}\cdot \nabla
\overline{\omega}=\nabla\cdot \biggl \lbrace D
\biggl\lbrack \nabla\overline{\omega}+{\beta(t)\over
C''(\overline{\omega})}\nabla\psi\biggr\rbrack\biggr\rbrace ,
\label{relax1}
\end{equation}
\begin{equation}
\beta(t)=-{\int D\nabla\overline{\omega}\cdot\nabla\psi d{\bf r}\over \int D{(\nabla\psi)^{2}\over C''(\overline{\omega})}d{\bf r}}, \qquad
\overline{\omega}=-\Delta\psi,
\label{relax2}
\end{equation}
where $\beta(t)$ is a Lagrange multiplier enforcing the energy
constraint $\dot E=0$ at any time. It is shown in \cite{pre} that
these equations increase monotonically the entropy ($H$-theorem $\dot
S\ge 0$) provided that $D>0$. Furthermore, a steady state of
(\ref{relax1}) is linearly dynamically stable iff it is a (local)
entropy maximum at fixed circulation and energy (minima or saddle
points of entropy are linearly unstable). Therefore, the relaxation
equations (\ref{relax1})-(\ref{relax2}) generically converge towards a
(local) entropy maximum (if there is no entropy maximum the solutions
of the relaxation equations can have a singular behaviour). If there
exists several local entropy maxima the selection will depend on a
complicated notion of {\it basin of attraction}. The diffusion coefficient
$D$ is not determined by the MEPP but it can be obtained from a
Taylor's type argument leading to $D=K\epsilon^{2}\omega_{2}^{1/2}$
where $\epsilon$ is the coarse-graining mesh size and $K$ is a
constant of order unity
\cite{physicaD}. Assuming that the relation (\ref{e6}) remains valid
out-of-equilibrium (see Appendix C of \cite{physicaD}), we get the
closed expression $D=K\epsilon^{2}/ \sqrt{C''(\overline{\omega})}$.
This position dependant diffusion coefficient, related to the strength
of the fluctuations, can ``freeze'' the system in a sub-region of
space (``bubble'') and account for {\it incomplete relaxation} and
lack of ergodicity \cite{rr,csr}. The relaxation equation
(\ref{relax1}) belongs to the class of nonlinear mean field
Fokker-Planck equations introduced in \cite{pre}. This relaxation
equation conserves only the robust constraints (circulation and
energy) and increases the generalized entropy (\ref{e3})-(\ref{e5})
fixed by the prior vorticity distribution $\chi(\sigma)$.  It differs
from the relaxation equations proposed by Robert \& Sommeria
\cite{rsmepp} for freely evolving flows which conserve all the
constraints of the 2D Euler equation ($E$, $\Gamma$ and all the
Casimirs) and monotonically increase the mixing entropy
(\ref{intro2}). In Eqs.  (\ref{relax1})-(\ref{relax2}), the
specification of the prior $\chi(\sigma)$ (determined by the
small-scale forcing) replaces the specification of the Casimirs
(determined by the initial conditions).  However, in both models, the
robust constraints $E$ and $\Gamma$ are treated microcanonically
(i.e. they are rigorously conserved). The relaxation equations of
Robert \& Sommeria
\cite{rsmepp} and Chavanis \cite{pre} are essentially
phenomenological in nature but they can serve as {\it numerical
algorithms} to compute maximum entropy states.  In that context, since
we are only interested by the stationary state (not by the dynamics),
we can take $D={\rm Cst.}$ and drop the advective term in the
relaxation equation. Then, Eq. (\ref{relax1}) can be used to construct
(i) arbitrary EHT statistical equilibria (ii) a subset of MRS
statistical equilibria (see the last paragraph of Sec. \ref{e}).


\section{\label{gamma}Explicit examples}

Let us consider, for illustration, the prior vorticity
distribution $\chi(\sigma)$ introduced by Ellis-Haven-Turkington
\cite{ellis} in their model of jovian vortices. It corresponds to a
de-centered Gamma distribution
\begin{equation}
\chi(\sigma)={1\over \Omega_2|\lambda|}R\biggl \lbrack {1\over \Omega_2\lambda}\biggl (\sigma+{1\over\lambda}\biggr );{1\over \Omega_2\lambda^{2}}\biggr \rbrack,
\label{g1}
\end{equation}
where $R(z;a)=\Gamma(a)^{-1}z^{a-1}e^{-z}$ for $z\ge 0$ and $R=0$
otherwise. The scaling of $\chi(\sigma)$ is chosen such that
$\langle\sigma\rangle=0$, ${\rm var}(\sigma)\equiv \langle\sigma^{2}\rangle
=\Omega_2$ and ${\rm skew}(\sigma)\equiv \langle\sigma^{3}\rangle /
\langle\sigma^{2}\rangle^{3/2}=2\Omega_2^{1/2}\lambda$. We get
\begin{equation}
Z(\Phi)=\hat\chi(\Phi)={e^{\Phi/\lambda}\over (1+\lambda \Omega_2
\Phi)^{1/(\Omega_2\lambda^{2})}}, \label{g2}
\end{equation}
\begin{equation}
\overline{\omega}(\Phi)=-(\ln \hat{\chi})'(\Phi)={-\Omega_2\Phi\over 1+\lambda \Omega_2 \Phi}.
\label{g3}
\end{equation}
Inversing the relation (\ref{g3}), we obtain
\begin{equation}
\label{g4}
-\Phi={1\over \Omega_2}{\overline{\omega}\over 1+\lambda\overline{\omega}}=C'(\overline{\omega}).
\end{equation}
After integration, we obtain the generalized entropy
\begin{equation}
\label{g5}
C(\overline{\omega})={1\over \lambda \Omega_2}\biggl\lbrack \overline{\omega}-{1\over\lambda}\ln (1+\lambda\overline{\omega})\biggr\rbrack.
\end{equation}
In the limit $\lambda\rightarrow 0$, the prior is the Gaussian distribution
\begin{equation}
\chi(\sigma)={1\over\sqrt{2\pi \Omega_2}}e^{-{\sigma^{2}\over 2\Omega_2}},
\label{g6}
\end{equation}
and we have $Z(\Phi)=e^{{1\over 2}\Omega_2\Phi^{2}}$,
$\overline{\omega}(\Phi)=-\Omega_2\Phi$,
$C(\overline{\omega})={\overline{\omega}^{2}\over 2\Omega_2}$.  The
generalized entropy $S=-\frac{1}{2\Omega_{2}}\int
\overline{\omega}^{2}d{\bf r}$ associated with a Gaussian prior is
proportional (with the opposite sign) to the coarse-grained
enstrophy: $S=-{\Gamma_{2}^{c.g.}}/({2\Omega_{2}})$ \cite{pre}. This
gaussian prior leads to Fofonoff flows \cite{fofonoff}  that have
oceanic applications.

When the prior is given by Eq. (\ref{g1}), the generalized entropy satisfies
$C''(\overline{\omega})=1/[\Omega_2(1+\lambda
\overline{\omega})^{2}]$ and we obtain a parametrization of the form
\begin{equation}
\label{g8}
{\partial \overline{\omega}\over\partial t}+{\bf u}\cdot \nabla \overline{\omega}=\nabla\cdot \biggl\lbrace D\biggl\lbrack \nabla\overline{\omega}+\beta(t)\Omega_2(1+\lambda \overline{\omega})^{2}\nabla\psi\biggr\rbrack\biggr\rbrace,
\end{equation}
\begin{equation}
\label{g9} \beta(t)=-{\int D\nabla{\omega}\cdot \nabla\psi d{\bf
r}\over \int D\Omega_2(1+\lambda
\overline{\omega})^{2}{(\nabla\psi)^{2}}d{\bf r}}, \
D={K\epsilon^{2}\Omega_2^{1/2}|1+\lambda  \overline{\omega}|}.
\end{equation}
For $\lambda=0$ (Gaussian limit), we get
\begin{equation}
\label{g11}
{\partial \overline{\omega}\over\partial t}+{\bf u}\cdot \nabla \overline{\omega}=\nabla\cdot \biggl\lbrace D\biggl\lbrack \nabla\overline{\omega}+\beta(t)\Omega_2\nabla\psi\biggr\rbrack\biggr\rbrace.
\end{equation}
\begin{equation}
\label{g12}
\beta(t)=-{\int D\nabla{\omega}\cdot \nabla\psi d{\bf r}\over \int D\Omega_2{(\nabla\psi)^{2}}d{\bf r}}, \qquad D={K\epsilon^{2}\Omega_2^{1/2}}.
\end{equation}
Since $D$ and $\Omega_{2}$ are uniform, we have
$D\overline{\omega}/Dt=D(\Delta\overline{\omega}-\beta(t)
\Omega_{2}\overline{\omega})$ with
$\beta(t)=-\Gamma_{2}^{c.g.}(t)/(2\Omega_{2} E)= S(t)/E$ (to arrive at
this result, we have used integration by parts in Eq. (\ref{g12})).


When the prior has two intense peaks $\chi(\sigma)=\delta(\sigma-\sigma_{0})+\delta(\sigma-\sigma_{1})$, the equilibrium coarse-grained vorticity is
\begin{equation}
\overline{\omega}=\sigma_{1}+{\sigma_{0}-\sigma_{1}\over 1+ e^{(\sigma_{0}-\sigma_{1})(\beta\psi+\alpha)}}.
\label{tp1}
\end{equation}
This is similar to the Fermi-Dirac statistics. Inverting this relation
to express $\Phi=\beta\psi+\alpha$ as a function of
$\overline{\omega}$ and integrating the resulting expression, we
obtain the generalized entropy
\begin{equation}
S[\overline{\omega}]=-\int \lbrack p\ln p+(1-p)\ln (1-p)\rbrack d{\bf r},
\label{tp2}
\end{equation}
where $\overline{\omega}=p\sigma_{0}+(1-p)\sigma_{1}$.  At
equilibrium, we have $\omega_{2}=1/C''(\overline{\omega})=(\sigma_{0}-\overline{\omega})(\overline{\omega}-\sigma_{1})$.
For the two-peaks distribution, we get a parametrization of the form
\begin{equation}
\label{tp4}
{\partial\overline{\omega}\over\partial t}+\overline{{\bf u}}\cdot \nabla\overline{\omega}=\nabla\cdot \left\lbrack D\left(\nabla\overline{\omega}+\beta(t)(\sigma_{0}-\overline{\omega})(\overline{\omega}-\sigma_{1})\nabla\psi\right )\right\rbrack.
\end{equation}
\begin{equation}
\label{tp5}
\beta (t)=-{\int D\nabla \overline{\omega}\cdot \nabla\psi d{\bf{r}}\over \int D (\sigma_{0}-\overline{\omega})(\overline{\omega}-\sigma_{1})(\nabla\psi)^{2}d{\bf{r}}}, \quad D=K\epsilon^{2}\omega_{2}^{1/2}.
\end{equation}
These are the same equations as in the MRS theory in the two levels
case $\omega\in \lbrace\sigma_{0},\sigma_{1}\rbrace$
\cite{miller,rs,lb,csr}. They amount to maximizing the
Fermi-Dirac-like entropy (\ref{tp2}) at fixed circulation and
energy. This entropy has been used by Bouchet \& Sommeria \cite{bs} to
model jovian vortices.  In the MRS viewpoint, this entropy describes
the free merging of a system with two levels of vorticity $\sigma_{0}$
and $\sigma_{1}$ while in the viewpoint developed here, it describes
the evolution of a forced system where the forcing has two intense
peaks described by the prior
$\chi(\sigma)=\delta(\sigma-\sigma_{0})+\delta(\sigma-\sigma_{1})$
\cite{physicaD}. Other examples of prior vorticity distributions and
associated generalized entropies are collected in \cite{pre}.

\section{Nonlinear dynamical stability}
\label{na}

Let us consider the Casimir functionals $S[\omega]=-\int
C(\omega)d{\bf r}$ where $C$ is any convex function ($C''>0$). Since
$S$, $E$ and $\Gamma$ are individually conserved by the 2D Euler equation, the
maximization problem
\begin{equation}
\label{na1}
\max_{\omega} \left\lbrace S[\omega] \ | \ E[\omega]=E,\Gamma[\omega]=\Gamma \right\rbrace,
\end{equation}
determines a steady state of the 2D Euler equation that is formally
nonlinearly dynamically stable \cite{ellis}. Writing the first
variations as $\delta S-\beta\delta E-\alpha\delta\Gamma=0$, the
steady state is characterized by a monotonic relation
$\omega=F(\beta\psi+\alpha)=f(\psi)$ where $F(x)=(C')^{-1}(-x)$. Let
us introduce the Legendre transform $J=S-\beta E$ and consider the
maximization problem
\begin{equation}
\label{na2} \max_{\omega} \left\lbrace J[\omega]=S[\omega]-\beta
E[\omega] \ | \Gamma[\omega]=\Gamma \right\rbrace.
\end{equation}
If we interpret $J$ as an energy-Casimir functional, the maximization
problem (\ref{na2}) corresponds to the Arnold criterion of formal
nonlinear dynamical stability. The variational problems (\ref{na1})
and (\ref{na2}) have the same critical points (cancelling the first
variations) but not necessarily the same maxima (regarding the second
variations). A solution of (\ref{na2}) is always a solution
of the more constrained problem (\ref{na1}). However, the reciprocal
is wrong. A solution of (\ref{na1}) is not necessarily a solution of
(\ref{na2}).  The maximization problem (\ref{na2}), and the associated
Arnold theorems, provide just a {\it sufficient} condition of
nonlinear dynamical stability. The criterion (\ref{na1}) of
Ellis-Haven-Turkington is more refined and allows to construct a
larger class of nonlinearly stable steady states. For example,
important equilibrium states in the weather layer of Jupiter are
nonlinearly dynamically stable according to the refined stability
criterion (\ref{na1}) while they do not satisfy the Arnold theorems
\cite{ellis}. The maximization problem (\ref{na2}) determines a
{\it subclass} of solutions of the maximization problem (\ref{na1}).
This is similar to a situation of ``ensemble inequivalence'' with
respect to the conjugate variables $(E,\beta)$ in thermodynamics
\cite{ineq,bb}. Indeed, (\ref{na1}) is similar to a criterion of
``microcanonical stability'' while (\ref{na2}) is similar to a
criterion of ``canonical stability'' in thermodynamics, where $S$ is
similar to an entropy and $J$ is similar to a free energy \cite{pre}. Canonical
stability implies microcanonical stability but the converse is wrong
in case of ensemble inequivalence \footnote{Since the EHT statistical
equilibria (with a given prior) satisfy a maximization problem of the
form (\ref{na1}) with $C(\overline{\omega})$ given by Eq. (\ref{e5}),
they are both thermodynamically stable (with respect to fine grained
perturbations $\delta\rho({\bf r},\sigma)$) and formally nonlinearly
dynamically stable (with respect to coarse-grained perturbations
$\delta\overline{\omega}({\bf r})$).  Note that the MRS statistical equilibria
may not satisfy the nonlinear dynamical stability criterion
(\ref{na1}) according to the discussion at the end of Sec.  \ref{e}.
This intriguing observation demands further investigation. }.
Since the relaxation equations (\ref{relax1})-(\ref{relax2}) solve the
maximization problem (\ref{na1}), they can serve as numerical
algorithms to compute nonlinearly dynamically stable stationary
solutions of the 2D Euler equation according to the criterion of
Ellis-Haven-Turkington. Note that if we fix $\beta$, the relaxation
equation (\ref{relax1}) increases monotonically the ``free energy''
$J=S-\beta E$ ($H$-theorem, $\dot J\ge 0$) until a (local) maximum of
$J$ at fixed $\Gamma$ is reached \cite{pre}. Therefore, we obtain a
numerical algorithm that solves the maximization problem (\ref{na2})
and determines a subclass of nonlinearly dynamically stable stationary
solutions of the 2D Euler equation corresponding to the  Arnold criterion.

\section{Conclusion}
\label{sec_conclusion}

In this paper, we have shown that the maximization of a functional
$S[\omega]$ at fixed circulation $\Gamma$ and energy $E$ in 2D
turbulence can have several interpretations. When $S$ is
given by (\ref{e3})-(\ref{e5}), this maximization problem determines:
(i) The whole class of stable EHT statistical equilibria for a given prior
vorticity distribution $\chi(\sigma)$ fixed by the small-scale
forcing. (ii) A subclass of stable MRS statistical equilibria for initial
conditions leading to a vorticity distribution $\chi(\sigma)$ at
statistical equilibrium. When $S$ is given by (\ref{e3}) where $C$ is
an arbitrary convex function, this maximization problem determines a
nonlinearly dynamically stable stationary solution of the 2D Euler
equation according to the refined EHT criterion. The next step is to
determine whether particular forms of generalized entropies are better
adapted than others to describe specific flows and whether they can be
regrouped in ``classes of equivalence'' \cite{pre}. For example, the
enstrophy functional turns out to be relevant for certain oceanic
situations
\cite{fofonoff} and the Fermi-Dirac-like entropy for jovian flows
\cite{bs}.  Working with a
suitable generalized entropy $S[\omega]$ with only two constraints
$(\Gamma, E)$ is more convenient than working with an infinite set of
Casimirs as in the MRS theory. This reduced maximization problem is
still very rich because, for any considered form of generalized
entropy $S[\omega]$, many bifurcations can take place in the parameter
space $(E,\Gamma)$
\cite{jfm2,ellis,bs}.

\appendix
\section{Generalized entropy}
\label{sec_ge}

We can introduce the generalized entropy $S[\overline{\omega}]$ in the
following manner. Initially, we want to determine the vorticity
distribution $\rho_*({\bf r},\sigma)$ which maximizes $S_{\chi}[\rho]$
with the robust constraints $E[\overline{\omega}]=E$,
$\Gamma[\overline{\omega}]=\Gamma$, and the normalization condition
$\int\rho\, d\sigma=1$. To solve this maximization problem, we can
proceed in two steps. {\it First step:} we determine the distribution
$\rho_1({\bf r},\sigma)$ which maximizes $S_{\chi}[\rho]$ with the
constraints $\int\rho\, d\sigma=1$ and a fixed vorticity profile
$\int\rho\sigma \, d\sigma =\overline{\omega}({\bf r})$ (note that
fixing $\overline{\omega}$ automatically determines $\Gamma$ and
$E$). This gives a distribution $\rho_1[\overline{\omega}({\bf
r}),\sigma]$ depending on $\overline{\omega}({\bf r})$ and
$\sigma$. Substituting this distribution in the functional
$S_{\chi}[\rho]$, we obtain a functional $S[\overline{\omega}]\equiv
S_{\chi}[\rho_1]$ of the vorticity $\overline{\omega}$. {\it Second
step:} we determine the vorticity field $\overline{\omega}_*({\bf r})$
which maximizes $S[\overline{\omega}]$ with the constraints
$E[\overline{\omega}]=E$ and
$\Gamma[\overline{\omega}]=\Gamma$. Finally, we have $\rho_*({\bf
r},\sigma)=\rho_1[\overline{\omega}_*({\bf r}),\sigma]$. Let us be more explicit. The distribution $\rho_1({\bf r},\sigma)$ that
extremizes $S_{\chi}[\rho]$ with the constraints $\int\rho\,
d\sigma=1$ and $\int\rho\sigma \, d\sigma =\overline{\omega}({\bf r})$
satisfies the first order variations $\delta S_{\chi}-\int \Phi({\bf
r}) \delta (\int \rho \sigma d\sigma) d{\bf r}-\int \zeta({\bf r})
\delta (\int \rho d\sigma) d{\bf r}=0$, where $\Phi({\bf r})$ and
$\zeta({\bf r})$ are Lagrange multipliers. This yields
\begin{eqnarray}
\rho_1({\bf r},\sigma)=\frac{1}{Z({\bf
r})}\chi(\sigma)e^{-\sigma\Phi({\bf r})},\label{ge1}
\end{eqnarray}
where $Z({\bf r})$ and $\Phi({\bf r})$ are determined by $Z({\bf
r})=\int \chi(\sigma)e^{-\sigma\Phi({\bf r})}d\sigma\equiv
\hat{\chi}(\Phi)$ and $\overline{\omega}({\bf r})=\frac{1}{Z({\bf
r})}\int \chi(\sigma)\sigma e^{-\sigma\Phi({\bf
r})}d\sigma=-(\ln\hat{\chi})'(\Phi)$.
This critical point is a {\it maximum} of $S_{\chi}$ with the
above-mentioned constraints since $\delta^{2} S_{\chi}=-\int
\frac{(\delta\rho)^{2}}{\rho} d{\bf r}d\sigma\le 0$. Then $S_{\chi}[\rho_1]=\int \rho_1(\sigma\Phi+\ln\hat\chi )\, d{\bf
r}d\sigma=\int (\overline{\omega}\Phi+\ln\hat{\chi}(\Phi))  \, d{\bf
r}$.
Therefore $S[\overline{\omega}]\equiv S_{\chi}[\rho_{1}]$ is given
by $S[\overline{\omega}]=-\int C(\overline{\omega}) \, d{\bf r}$
with $C(\overline{\omega})=
-\overline{\omega}\Phi-\ln\hat{\chi}(\Phi)$.
Now, $\Phi({\bf r})$ is related to $\overline{\omega}(\bf{r})$ by $\overline{\omega}({\bf r})=-(\ln\hat{\chi})'(\Phi)$. This implies that $C'(\overline{\omega})=-\Phi=-[(\ln\hat\chi)']^{-1}(-\overline{\omega})$
so that
\begin{eqnarray}
C(\overline{\omega})=-\int^{\overline{\omega}}
[(\ln\hat\chi)']^{-1}(-x)dx.\label{ge8}
\end{eqnarray}
This is precisely the generalized entropy (\ref{e5}). Therefore, $\rho_{*}({\bf r},\sigma)=\rho_1[\overline{\omega}_*({\bf r}),\sigma]$ is a maximum of $S_{\chi}[\rho]$ at fixed $E$ and $\Gamma$ iff $\overline{\omega}_{*}({\bf r})$ is a maximum of $S[\overline{\omega}]$ at fixed  $E$ and $\Gamma$.






\end{document}